\documentclass[aps,twocolumn,prl,floatfix]{revtex4-1}

\usepackage{graphicx}
\usepackage{amssymb}
\usepackage{amsmath}

\usepackage{color}

\usepackage[normalem]{ulem}

\newcommand {\be}{\begin{equation}}
\newcommand {\ee}{\end{equation}}
\newcommand {\ba}{\begin{eqnarray}}
\newcommand {\ea}{\end{eqnarray}}

\begin{document}
\title{Entangled Dynamics in Macroscopic Quantum Tunneling of Bose-Einstein Condensates}
\author{Diego A. Alcala, Joseph A. Glick, and Lincoln D. Carr}
\affiliation{Department of Physics, Colorado School of Mines, Golden, CO 80401, USA}
\date{\today}

\begin{abstract}
Tunneling of a quasibound state is a non-smooth process in the entangled many-body case.  Using time-evolving block decimation, we show that repulsive (attractive) interactions speed up (slow down) tunneling, which occurs in bursts.  While the escape time scales exponentially with small interactions, the maximization time of the von Neumann entanglement entropy between the remaining quasibound and escaped atoms scales quadratically. Stronger interactions require higher order corrections. Entanglement entropy is maximized when about half the atoms have escaped.

\end{abstract}
\maketitle

Tunneling is one of the most pervasive concepts in quantum mechanics and is essential to contexts as diverse as $\alpha$-decay of nuclei~\cite{Gurney1929}, vacuum states in quantum cosmology~\cite{coleman_fate_1977} and chromodynamics~\cite{ostrovsky_forced_2002}, and photosynthesis~\cite{Collini2010}. Macroscopic quantum tunneling (MQT), the aggregate tunneling behavior of a quantum many-body wavefunction, has been demonstrated in many condensed matter systems~\cite{Thomas1996,Nakamura1999} and is one of the remarkable features of Bose-Einstein Condensates (BECs), ranging from Landau-Zener tunneling in tilted optical lattices~\cite{anderson1998} to the AC and DC Josephson effects in double wells~\cite{albiez_direct_2005,steinhauer2007}, as well as their quantum entangled generalizations~\cite{Beinke2015}.  The original vision of quantum tunneling was in fact the \textit{quantum escape} or quasibound problem by Gurney and Condon in 1929~\cite{Gurney1929}, and recently the first mean-field or semiclassical observation of quantum escape has been made in Toronto~\cite{carr2016a}.  However, with the rise of entanglement as a key perspective on quantum many-body physics, the advent of powerful entangled dynamics matrix-product-state (MPS) methods~\cite{schollwoeck2011,openMPS}, and the possibility of observing the moment-to-moment time evolution of quasibound tunneling dynamics directly in the laboratory~\cite{stadler2012,Zurn2013,Kaufman2014,Schweigler2015,carr2016a} it is the right time to revisit quantum escape.  In this Letter, we take advantage of the powerful new toolset for quantum many-body simulations~\cite{AlPS,openMPS} to show that the many-body quantum tunneling problem differs in key respects from our expectations from semiclassical and other well-established approaches to tunneling.

Specifically, we use time-evolving block decimation (TEBD) to follow lowly entangled matrix product states~\cite{vidal2003,schollwoeck2011} for the quantum escape of a quasibound ultracold Bose gas initially confined behind a potential barrier.  Our use of a Bose-Hubbard Hamiltonian~\cite{bloch2008} can be viewed either as a discretization scheme or as an explicitly enforced optical lattice used to control the tunneling dynamics.  Unlike instanton and semiclassical approaches, we are able to follow the von Neumann entanglement entropy, number fluctuations, quantum depletion, and other quantum many-body aspects of time evolution of the many-body wavefunction.  Such measures clarify when semiclassical approaches are and are not applicable.  They also show that hiding in the semiclassical averaged picture are other many-body features with radically different scalings: the \emph{escape time} $t_{\mathrm{esc}}$, i.e., the time at which the average number of remaining quasibound atoms falls to $1/e$ of its initial value, increases (decreases) as an exponential with attractive (repulsive) interactions for a limited range of interactions near zero. We will show that in order to accurately describe the scaling of $t_{\mathrm{esc}}$ and other many-body observables over many interaction strengths, one must include the effect of higher corrections.

Whether between discrete states in a double well~\cite{rhaghavan1999,albiez_direct_2005}, in Landau-Zener~\cite{cristianiM2002} and orbital angular momentum contexts~\cite{2013NJPh...15j3006H}, for quantum escape~\cite{carr2004b,carr_macroscopic_2005}, or even in variational parameter space~\cite{ueda_macroscopic_1998}, MQT has up till now mainly been treated under semiclassical approximations such as the instanton approximation and JWKB, as well as the nonlinear Schrodinger equation (NLS).  The NLS approach already establishes non-smooth time-evolution of a quasibound state in the form of ``blips'' or bursts of condensate~\cite{dekel_dynamics_2010}, although mean-field theory sometimes gives incorrect predictions in this regard; we demonstrate that the burst predictions are correct.  Beyond mean-field, semiclassical, and instanton approaches, two time-evolving many-body studies have been performed recently. First, an explicit comparison between instanton and TEBD Bose-Hubbard based predictions has been performed for superfluid decay~\cite{danshita2010,danshita2012}, establishing explicit numerical limits on the instanton approach; this method is nearly identical to ours but treats discrete-to-discrete state or double-well type tunneling, in this case between two rotational states on a ring.  Second, the quantum escape problem has been studied with the first-quantization-based time-adaptive many-body method known as multi-configurational Hartree-Fock theory~\cite{lode2012,Beinke2015}; this work treated quantum depletion but not von Neumann entropy and number fluctuations.  In contrast, our approach accesses a wide variety of quantum measures to elucidate the underlying many-body quantum features of quasibound escape dynamics, and shows the explicit convergence to mean-field type dynamics.

Consider a system of $N$ bosons at zero temperature in the canonical ensemble.  To simulate such a system, we can either invoke an explicit optical lattice of $L$ sites, deep enough for tight binding and single band approximations to be valid; or we can simply choose a discretization scheme.  Either way the Bose Hubbard Hamiltonian (BHH) is an appropriate model:
\begin{equation}
\label{eq:BHH}
\hat{H} = -J\sum_{i=1}^{L-1}(\hat{b}_{i+1}^\dagger\hat{b}_i+\mathrm{h.c.})+\sum_{i=1}^L [\frac{U}{2}\hat{n}_i(\hat{n}_i-\hat{1})+V^{\mathrm{ext}}_i \hat{n}_{i}].
\end{equation}
In Eq.~\eqref{eq:BHH}, $J$ is the energy of hopping and $U$ determines the on-site two-particle interactions. An external rectangular potential barrier, of width $w$ and height $h$, is given by $V_i^{\mathrm{ext}}$. The field operator $\hat{b}_{i}^\dagger$ ($\hat{b}_{i}$) creates (annihilates) a boson at the $i\mathrm{th}$ site and $\hat{n}_{i} \equiv \hat{b}_{i}^\dagger \hat{b}_{i}$. We will work in hopping units: energies are scaled to $J$ and time $t$ to $\hbar/J$. We use open boundary conditions, as convenient for TEBD.   TEBD is a matrix product state numerical method that time evolves Eq.~\eqref{eq:BHH} on a time-adaptive reduced Hilbert space, given that the system is lowly entangled. TEBD is a superior method because it gives us access to quintessential many-body quantities like entanglement. Instanton methods offer another approach towards calculating tunneling rates within a semiclassical approximation~\cite{vainshtein_abc_1982}, but are rapidly rendered inaccurate for larger interaction strengths~\cite{danshita_accurate_2010}, whereas TEBD suffers from no such limitations.

To describe the system from a mean field perspective, the discrete NLS (DNLS) may either be obtained via discretization of the NLS or from a mean field approximation of the BHH. In the latter case, one can propagate the field operator $\hat{b}_{i}$ forward in time using the BHH in the Heisenberg picture:  $i \hbar \partial{}_t \hat{b}_{i}=[\hat{b}_{i},\hat{H}]$. Assuming the many-body state is a product of Glauber coherent states, $\langle \hat{b}_i^{\dagger}\hat{b}_i\hat{b}_i\rangle = \psi_i^*\psi_i\psi_i$, where $\psi_i\equiv\langle \hat{b}_{i} \rangle$, leads to the DNLS:
\begin{equation}
\label{eq:DNLS}
\textstyle i\hbar\dot{\psi_i}=-J(\psi_{i+1}+\psi_{i-1}) + g|\psi_i|^2 \psi_i+V^{\mathrm{ext}}_{i} \psi_i.
\end{equation}
In Eq.~\eqref{eq:DNLS}, the condensate order parameter, $\psi_i$, is normalized to the number of atoms, $N = \sum^{L}_{i=1} |\psi_i|^2$. Mean field simulations are performed using a fourth-order Runge-Kutta adaptation of Eq.~\eqref{eq:DNLS}.  The BHH approaches the DNLS in the mean field limit $N\to \infty$, $U \to 0$, $NU/J=\mathrm{const.}$  We emphasize that both the BHH and the DNLS are single band models, valid when the many-body wavefunction covers many sites and has variations larger than the lattice constant. A true continuum limit is possible for $N J / L = \mathrm{const.}$, $N/L \to 0$ and $J \to \infty$; however, this would restrict us numerically to very small numbers of atoms~\cite{muth_fermionization_2010} and prevent us from approaching the mean field limit of $NU=\mathrm{const.}$, $N\to \infty$, $U\to 0$; it can also require different discretization schemes than the BHH depending on the interaction strength and regime of interest.  We therefore restrict ourselves to the semi-discrete regime appropriate to both the BHH and DNLS.

\begin{figure}[t]
\begin{center}
\includegraphics[scale=0.36]{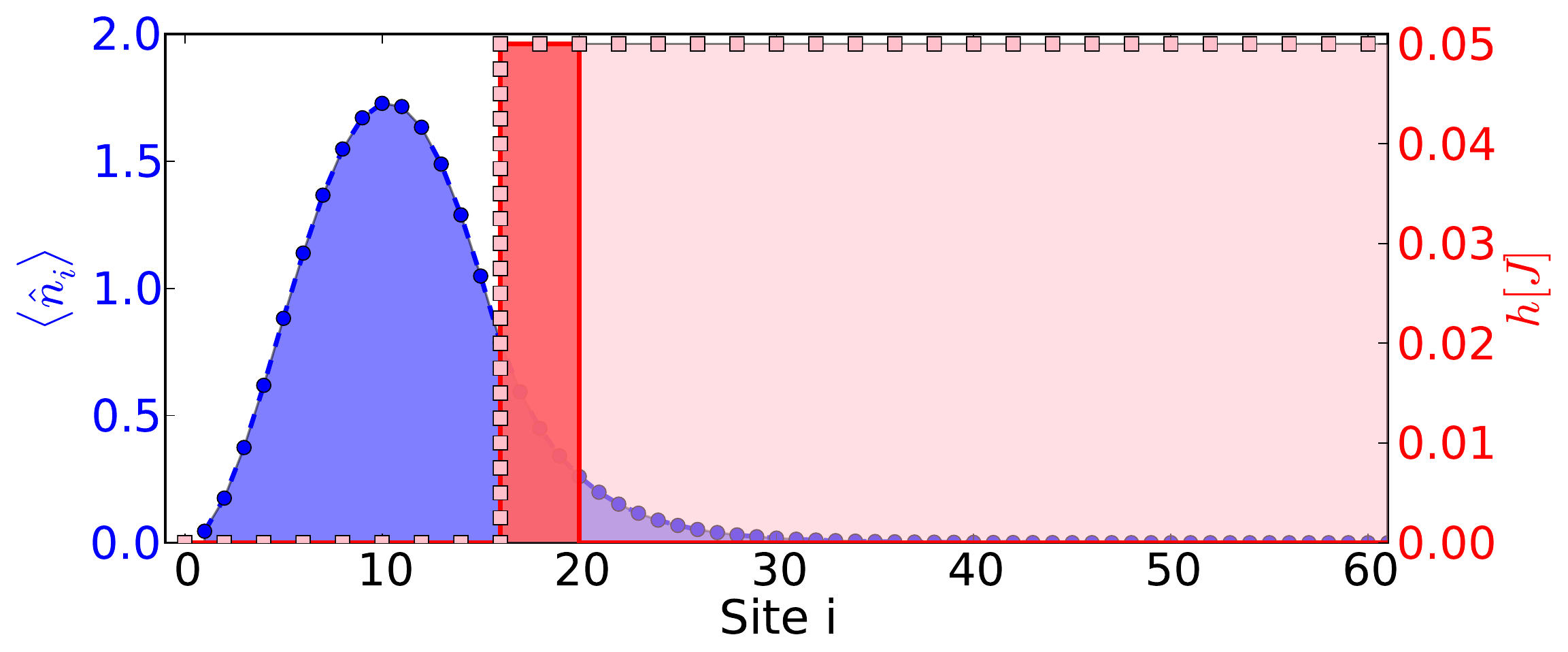}
\caption{\label{fig:InitialState} \emph{Initial Quasibound State.}  The many-body wavefunction for $N=20$ with $NU/J=+0.15$ (blue shaded region, points show actual TEBD results for the density average $\langle \hat{n}_i \rangle$j) is first localized to the left behind the barrier (red line, red and pink shaded areas) via relaxation in imaginary time with a barrier of height $h$ and initial width $w_{I}$.  At $t=0$ in real time propagation the barrier is reduced to width $w$ (solid red line, red shaded area) so the now quasibound Bose gas can commence macroscopic quantum tunneling.  The hard wall at the left and relatively small barrier area pushes the density tail to partially extend to the right.}
\end{center}
\end{figure}
\begin{figure}
\begin{center}
\includegraphics[scale=0.32]{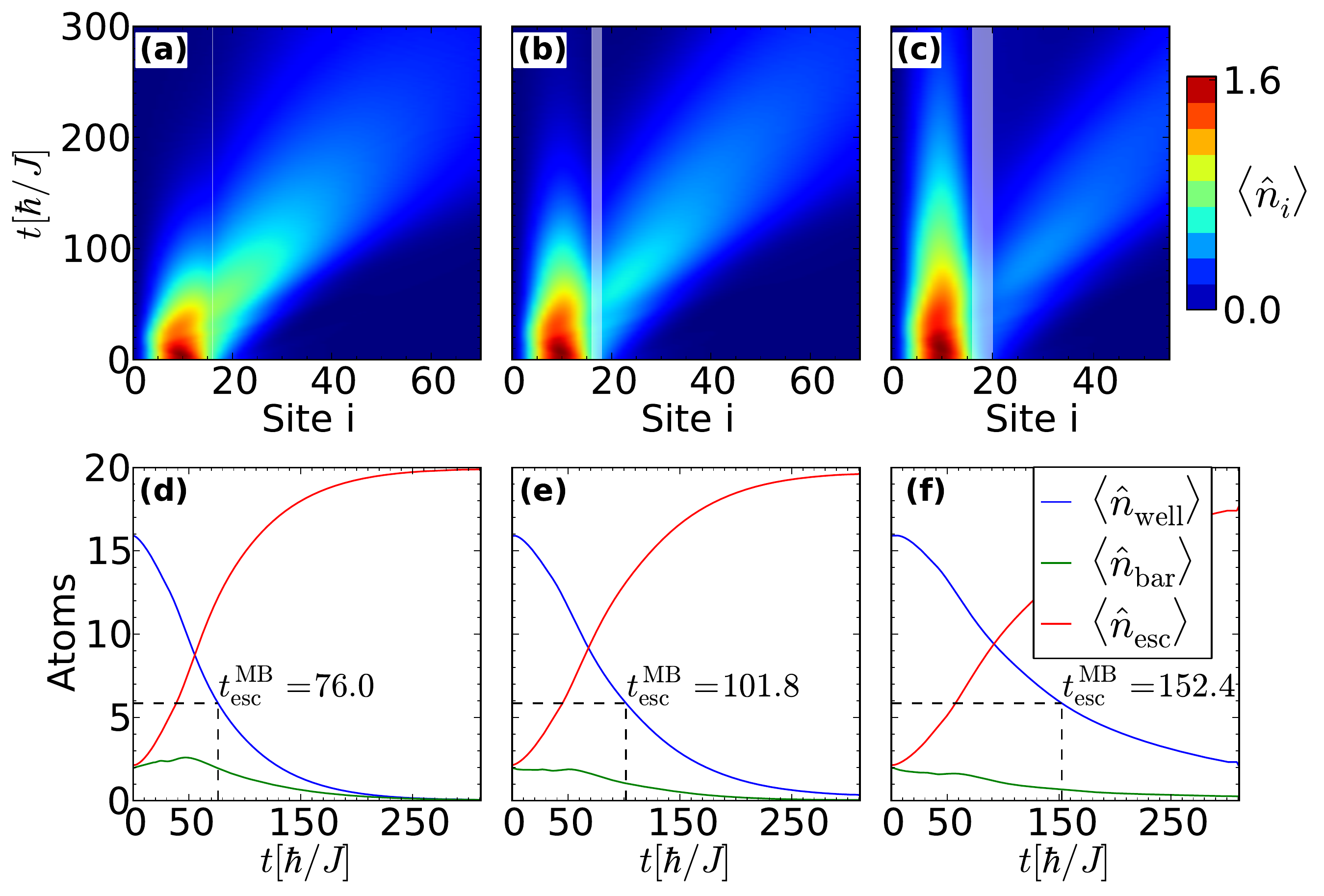}
\caption{\label{fig:AvgNumChangeW} \emph{Many-Body Tunneling, and Calculation of Decay Time.} Barrier widths (a,d) $w=1$, (b,e) $w=3$ , and (c,f) $w=5$. Top row: average atom number per site. Bottom row: number in well $\langle \hat{n}_{\mathrm{well}} \rangle$ (blue), number in barrier $\langle \hat{n}_{\mathrm{bar}} \rangle$ (green), and escaped $\langle \hat{n}_{\mathrm{esc}} \rangle$ (red) atoms;  the $1/e$ decay time all $\pm 0.1$. All plots for $NU/J=+0.30$ with $N=20$.}
\end{center}
\end{figure}

We initialize the many-body wavefunction via imaginary time relaxation to trap the atoms in a quasibound state behind the barrier as illustrated in Fig.~\ref{fig:InitialState}. We set $V^{\mathrm{ext}}$ to height $h=0.05$ and width $w_{I}$, effectively reducing the system size. At $t=0$, in real time, the barrier is decreased to width $w$, where $w$ is typically one to five sites, such that the atoms can escape on a time scale within reach of TEBD simulations. We choose $L$ large enough so that reflections from the box boundary at the far right do not return to the barrier in simulation times of interest: $t_\mathrm{reflect}  \gg  t_{\mathrm{esc}}$.  Evolving in real time, we first make a coarse observation of the dynamics of MQT in Fig.~\ref{fig:AvgNumChangeW} by plotting the average atom number in different regions for repulsive interactions, in order to determine $t_\mathrm{esc}$.  We find similar results for attractive interactions, but with larger $t_{\mathrm{esc}}$.

\begin{figure}[t]
\begin{center}
\includegraphics[scale=0.31]{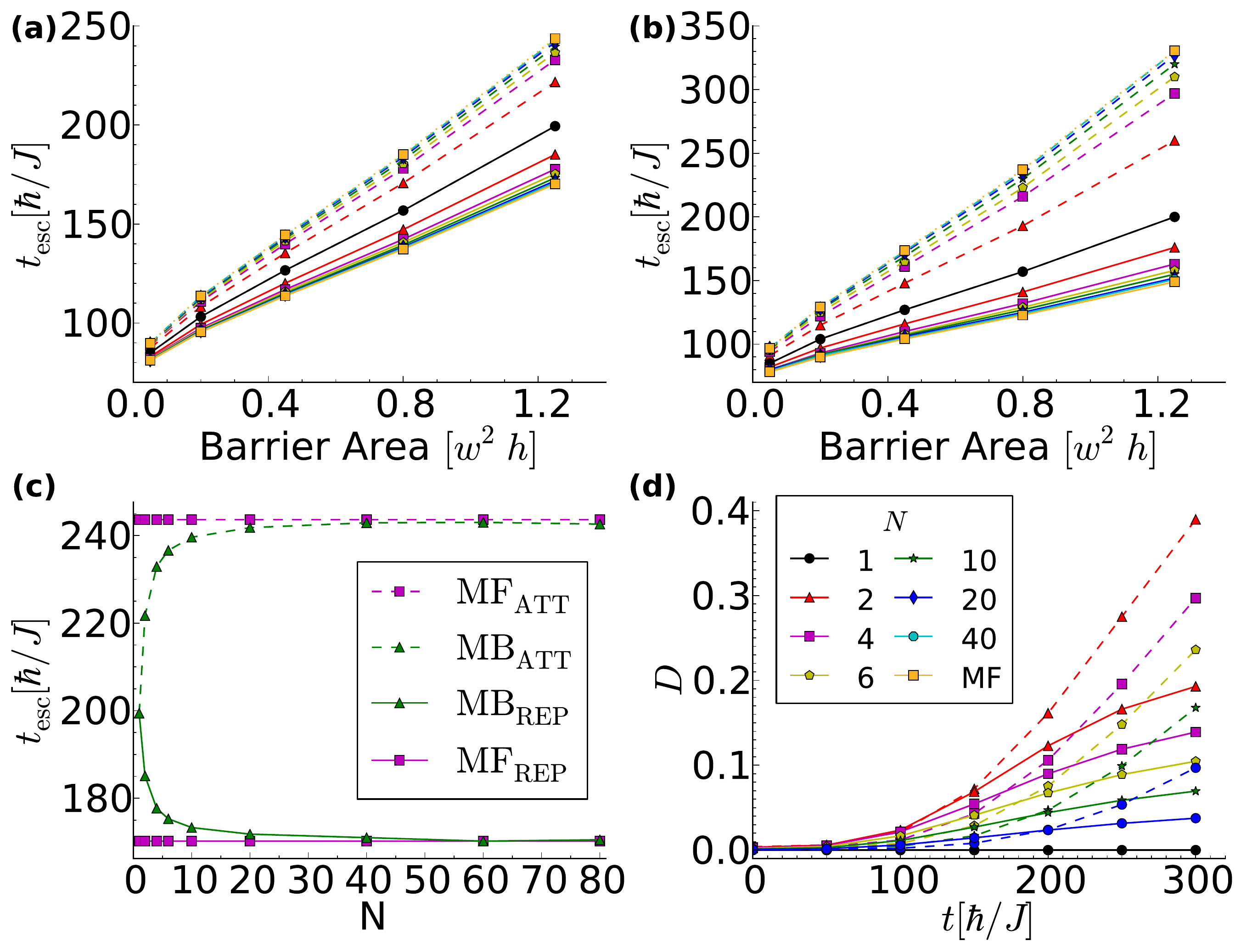}
\caption{\label{fig:NUovJEscapeSummary} \emph{Many Body (MB) vs. Mean Field (MF) Escape Time Predictions.} Solid lines: Repulsive (REP). Dashed lines: Attractive (ATT). (a)-(b) Dependence of $t_{\mathrm{esc}}^{\mathrm{MB}}$ on barrier area and atom number for (a) $N U/J = \pm 0.15$ and (b) $NU/J=\pm 0.30$. (c) $t_{\mathrm{esc}}^{\mathrm{MB}}$ plateaus towards $t_{\mathrm{esc}}^{\mathrm{MF}}$ for $10$ to $80$ atoms as shown for $NU/J = \pm 0.15$ and $w=5$. (d) Fragmentation for $NU/J=\pm 0.30$ and $w=5$. Curves are a guide to the eye, points represent actual data with error bars smaller than data point in all panels. Panel (d) legend corresponds to (a),(b), and (d).
}
\end{center}
\end{figure}

How do many-body predictions compare to mean field ones?  We define $t_{\mathrm{esc}}^{\mathrm{MF}}$ and $t_{\mathrm{esc}}^{\mathrm{MB}}$ as the mean field and many-body escape times, respectively.  For fixed $N U/J$, $w$, and $h$, the DNLS gives the same result independent of $N$ and $U$; $t_{\mathrm{esc}}^{\mathrm{MB}} \to t_{\mathrm{esc}}^{\mathrm{MF}}$ only in the large $N$ small $|U|$ mean field limit; and $w^2 h$ determines the barrier area. Figure~\ref{fig:NUovJEscapeSummary} illustrates our exploration of this parameter space.
The dynamics of MQT predicted by the DNLS and BHH differ strongly when $N$ is small. Generally, the DNLS predicts $t_{\mathrm{esc}}^{\mathrm{MB}}$ well when $N$ is sufficiently large. For example, in Fig.~\ref{fig:NUovJEscapeSummary}(c) for repulsive (attractive) interactions $NU/ J = +0.15$ ($NU/ J = -0.15$) and barrier width $w=5$, the BHH predicts a decrease (increase) in $t_{\mathrm{esc}}^{\mathrm{MB}}$, approaching a nearly constant value for $N \gtrsim 20$. This same trend is apparent for various barrier areas, see Fig.~\ref{fig:NUovJEscapeSummary}(a,b). In Fig.~\ref{fig:NUovJEscapeSummary}(d) we also show the quantum depletion $D$ or fragmentation, for $N U/J=\pm0.30$, $w=5$, $D \equiv 1-(\lambda_1) / (\sum_{m=1}^{L}\lambda_m)$ where $\{\lambda_m\}$ are the eigenvalues of the single particle density matrix $\langle \hat{b}_i^{\dagger}\hat{b}_j\rangle$, and $\lambda_1$ is the largest eigenvalue; larger $D$ corresponds to a more fragmented (less condensed) state. The largest fragmentation for both attractive and repulsive interactions occurs for $N=2$. As $N$ increases, depletion decreases monotonically, with $N=20$ reaching $D \approx 0.10$ ($D \approx 0.04$) for attractive (repulsive) interactions. This decreased fragmentation allows the DNLS to give accurate predictions for $t_{\mathrm{esc}}^{\mathrm{MB}}$ for larger $N$.

Systematic error in TEBD~\cite{vidal_efficient_2004} for $t_{\mathrm{esc}}^{\mathrm{MB}}$ results from the Schmidt truncation  $(\chi)$, the truncation in the on-site Hilbert space dimension $(d)$, and the time resolution at which we write out data $(\delta t)$.  The hardest many-body measures to converge, such as the block entropy, at $\chi=35$ have an error $\lesssim 10^{-3}$ for $N=70$, and were checked up through $\chi=55$; due to small $U$ and effective system size, much lower $\chi$ is required than usual in TEBD.  For up to $N=10$ we have not truncated $d$, but for larger $N$ up to 80, we truncated attractive (repulsive) to $d=20$ ($d=15$).  A lower truncation results in decreased $t_{\mathrm{esc}}^{\mathrm{MB}}$, e.g. by 10\% for $d=5$, $NU/J=-0.1$, and $N=10$, even though $\mathrm{max}(\langle\hat{n}\rangle) < 1$, since more weight is given to spread-out Fock states. The attractive BHH requires much higher $d$ than the repulsive BHH, since $U<0$ increases number fluctuations in high density regions, i.e., behind the barrier at $t=0$.  In both cases, in general we find on-site number fluctuations play a surprisingly strong role in tunneling processes compared to usual for TEBD.  The BHH also has a number of sources of systematic error, the most important of which is virtual fluctuations to the second band; however, since we compare single-band DNLS to single-band BHH this does not effect our comparison.  In general we expect fluctuations to higher bands will speed up tunneling; therefore out calculations may be taken as a lower bound for experiments.

\begin{figure}[t]
\begin{center}
\includegraphics[scale=0.31]{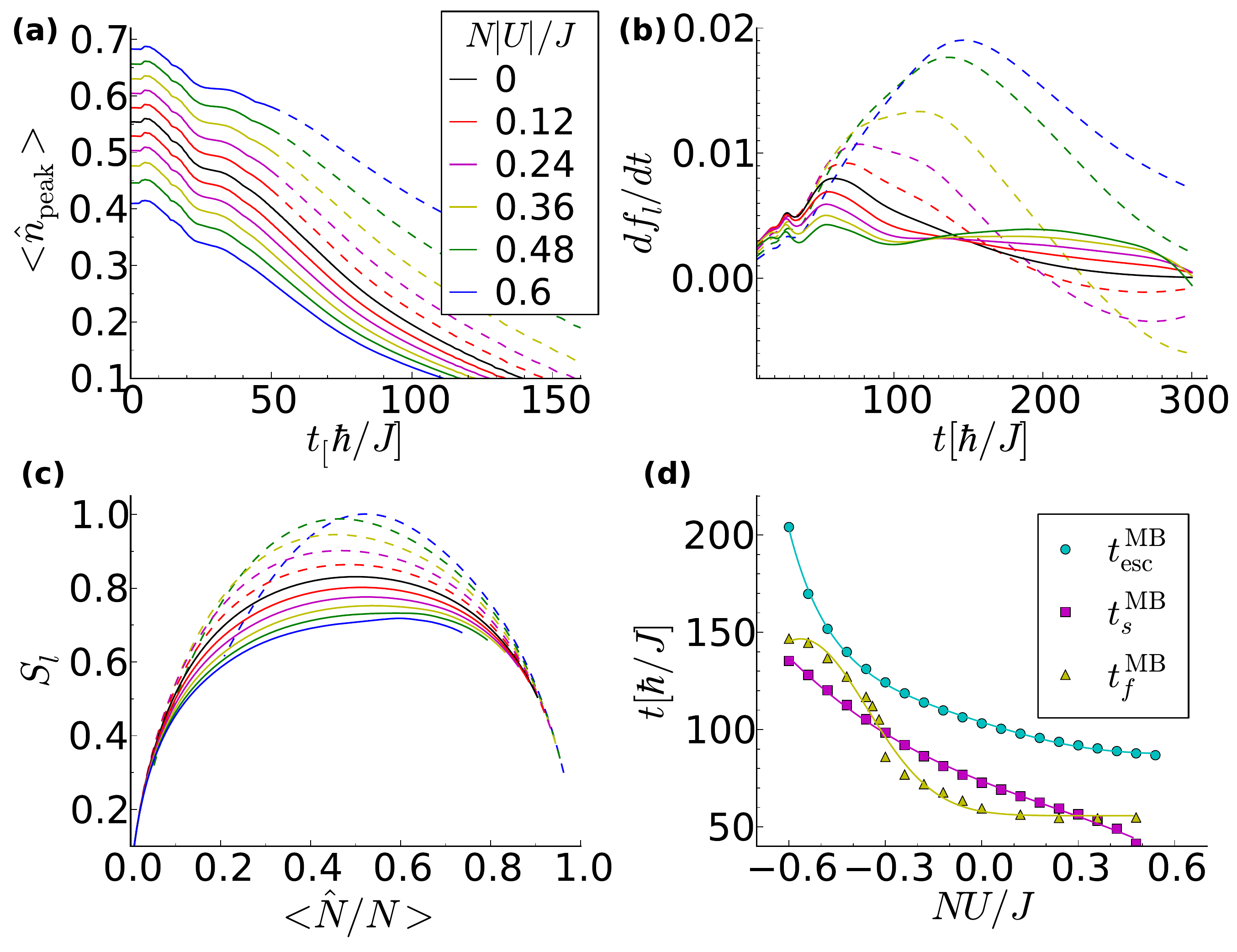}
\caption{\label{fig:peakFluctEnt}
\emph{Many-body Quantum Measures.} Solid lines: Repulsive. Dashed lines: Attractive. (a) Average number at the density peak shows bursts of atoms~\cite{dekel_dynamics_2010}. Early time attractive lines solid to make dynamics distinguishable.   (b) Time derivative of number fluctuations in the number of trapped atoms are smaller for repulsive interactions. (c) Nearly universal curve for the entropy of entanglement vs. the average number of trapped atoms. (d) Observables demonstrate very different scaling with interaction. Points show actual data (error bars smaller than points), while lines are best fit curves. All plots treat $N=6$. Panel (a) legend correspond to (a),(b), and (c).}
\end{center}
\end{figure}
In Fig.~\ref{fig:peakFluctEnt}(a) we plot the average number at the peak of the many-body wavefunction. There are points in time when the number density exhibits quadratic decay, and others during which it is nearly constant, similar to the density bursts found by Dekel \textit{et al.}~\cite{dekel_dynamics_2010}; thus their predictions are correct even in the many-body regime. The first burst is nearly independent of $U$. The initial flat horizontal region and burst originate from the wave-function pushing away from the leftmost infinite boundary and interacting with the barrier. For attractive interactions, the initial small increase in $\hat{n}_{\mathrm{peak}}$ is due to the atoms fluctuating towards the peak, attracted by the strong concentration of atoms. For repulsive interactions this increase occurs because the wavefunction collides with the barrier after pushing away from the infinite wall, causing a slight swell in $\hat{n}_{\mathrm{peak}}$.  All subsequent dynamics appear to be dependent on $U$.

To characterize the quantum nature of MQT, in Fig.~\ref{fig:peakFluctEnt}(b) we plot the time derivative of fluctuations in the number of atoms behind the barrier $df_{l}/dt$, where $f_l=(\langle N^2_l \rangle - \langle N_l \rangle^2) / \langle N_l \rangle$, $N_l$ is the number of atoms to the left of site $l$, and $l$ is taken at the outer edge of the barrier. Once MQT commences, the maximum value of $df_{l}/dt$ in time increases with decreasing $U$ because number densities just outside the barrier have more influence to ``pull'' additional atoms through the barrier for attractive interactions.  Repulsive interactions, in comparison, suppress tunneling, so $df_{l}/dt$ does not increase as much.

Of particular interest to MQT is the von Neumann block entropy characterizing entanglement between the remaining quasibound atoms and the escaped atoms, $S_l\equiv -\mathrm{Tr}(\hat{\rho}_l \log \hat{\rho}_l)$, where $\hat{\rho}_l$ is the reduced density matrix for the well plus barrier.  The key features of $S_l$ are illustrated in a nearly universal curve in Fig.~\ref{fig:peakFluctEnt}(c): on the lower right side tunneling has not yet commenced. $S_l$ maximizes part way through the tunneling process in the center of the curve, at $N_l/N \simeq 1/2$; and $S_l$ then decreases again to the left as the atoms finish tunneling out.

Define $t_{s}$ as the time at which $S_{l}$ is maximized and define $t_{f}$ as the time at which the slope of the number fluctuations ($df_l/dt$) is largest before $t_{f}$. We find $t_{s}$, $t_{f}$, and $t_\mathrm{esc}^{\mathrm{MB}}$ increase with decreasing $U$, as shown in Fig.~\ref{fig:peakFluctEnt}(d). As $NU/J$ decreases, we approach the self-trapping regime, where escape times become much longer than the lifetime of the system.  While $t_{\mathrm{esc}}$ increases smoothly as $NU/J$ decreases, $df_{l}/dt$ is strongly influenced by change in $NU/J$, with a noticeable increase near $NU/J \approx -0.3$, and a steady flattening-out as we approach self-trapping interaction strength.  A best fit line for $t_{f}$ covering all $NU/J$ requires an exponential of a second order polynomial, while an exponential fits well for $-0.3 < NU/J < 0.3$, as also found in Ref.~\cite{Kolovsky2010} for bright solitons tunneling in a tilted optical lattice.  In the coarser measure $t_{\mathrm{esc}}^{\mathrm{MB}}$, we find exponential scaling when $-0.4 < NU/J < 0.4$. In order to accurately capture the strong interaction regimes $N|U|/J \gtrsim 0.4$, we need a third order polynomial in the exponential, as shown in the fit in Fig.~\ref{fig:peakFluctEnt}(d). We find that $t_{s}$ scales linearly only for $-0.1 < NU/J < 0.1$, quadratically for $-0.4 < NU/J < 0.4$, and requires a cubic polynomial fit to cover the entire interaction regime.  Results in Fig.~\ref{fig:peakFluctEnt} are for $N=6$; we found similar results for up to $N=20$, although simulations are limited in the large $|U|$ regime.

\begin{figure}[t]
\begin{center}
\includegraphics[scale=0.4]{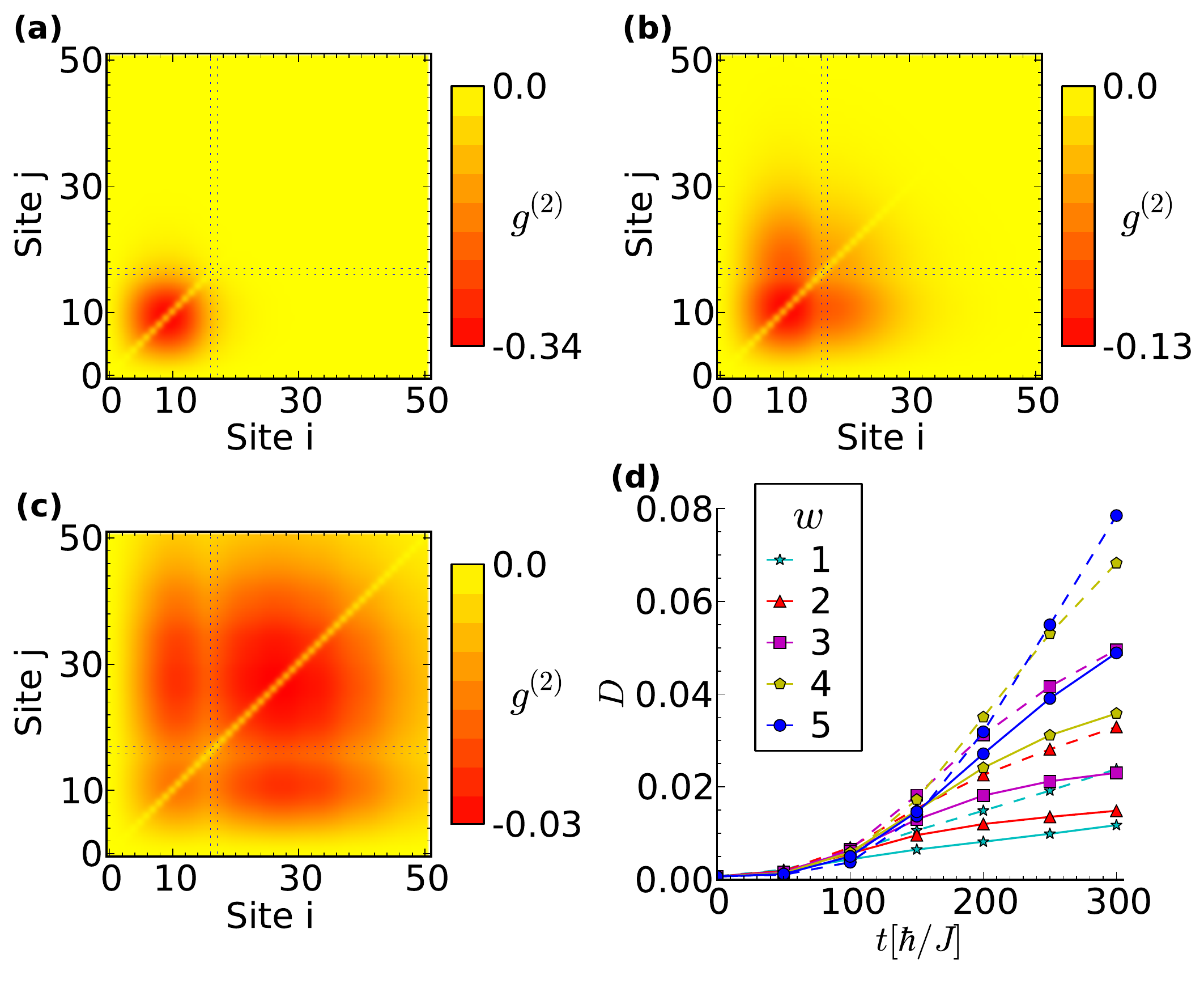}
\caption{\label{fig:geetwo} \emph{Time-dependence of Density-Density Correlations.}  (a)-(c) $g^{(2)}$ shows correlations between trapped and escaped atoms. The barrier, indicated by dotted lines, breaks up negatively-correlated regions (red); shown are time slices at (a) $t=0$, (b) $t=62 \approx t_{s}$, and (c) $t=125 \approx t_{\mathrm{esc}}^{\mathrm{MB}}$. (d) Quantum depletion grows rapidly for $N=2$ with $NU/J=\pm 0.15$. Solid lines: Repulsive. Dashed lines: Attractive. Curves are a guide to the eye, points represent actual data (error bars smaller than points). }
\end{center}
\end{figure}

Another experimental signature is the density-density correlations, $g^{(2)}_{ij}=\langle \hat{n}_i \hat{n}_j \rangle - \langle \hat{n}_i \rangle \langle \hat{n}_j \rangle$, extractable from noise measurements~\cite{altman_probing_2004,greiner_probing_2005}; $g^{(2)}$ is zero in mean field theory.  As customary, we subtract off the large diagonal matrix elements of $g^{(2)}$ to view the underlying off-diagonal structure.  In Fig.~\ref{fig:geetwo}(a)-(c) we show $g^{(2)}$ for $N=40$, $NU/J=-0.015$, and $w=2$, dividing up the system to observe correlations between the three physical regions: trapped, under the barrier, and escaped.  We initially observe near-zero correlations everywhere except near the many-body wavefunction peak. At $t=62 \approx t_{s}$, $g^{(2)}$ shows many negatively-correlated regions ($g^{(2)}<0$) which are broken up by the potential barrier. In Fig.~\ref{fig:geetwo}(d) we also show quantum depletion $D$ for $NU/J=\pm 0.15$ with $N=2$ and $w=1,2,3,4,5$.
$D$ increases with increasing $w$. In comparison to Fig.~\ref{fig:NUovJEscapeSummary}(d) ($NU/J=\pm 0.30$), $D$ doesn't become as large for Fig.~\ref{fig:geetwo}(d) ($NU/J=\pm 0.15$) because of the smaller $N|U|/J$ value.  The growth in $D$ emphasizes the many-body nature of the escape process.

In conclusion, we have performed quantum many-body simulations of the macroscopic quantum tunneling of attractive and repulsive bosons using TEBD to time-evolve the Bose-Hubbard Hamiltonian, treating the original 1929 quasibound or quantum escape problem.  We found strong deviations from mean field predictions and provided quantitative boundaries by which one can judge the legitimacy of applying mean field theory to this problem. Even a low average order moment like escape time was shown to deviate from simple exponential scaling for strong interactions. Higher order quantum measures like entropy of entanglement between the quasibound and escaped atoms, and the slope of number fluctuations, reached a maximum at times which exhibited scaling behaviors with interactions ranging from polynomial to exponential to exponential of a polynomial, showing tunneling dynamics are far richer in the quantum many body picture.  We showed the many-body extension to the predictions of Dekel \textit{et al.} regarding number density bursts~\cite{dekel_dynamics_2010}. Finally, our study shows that many-body effects in macroscopic quantum tunneling can be experimentally observed via number fluctuations and density-density correlations as well as dependence of escape time on interactions.

We thank Veronica Ahufinger, Jen Glick, Mark Lusk, Kenji Maeda, Marie Mclain, Shreyas Potnis, Anna Sanpera, Aephraim Steinberg, Marc Valdez, Michael Wall, David Wood, and Xinxin Zhao for valuable discussions. This work was supported by NSF and the AFOSR.



%

\end{document}